\documentclass[twocolumn,reprint,showpacs, superscriptaddress, amsmath,amssymb, aps, pra]{revtex4-1}
\usepackage{graphicx}
\usepackage{amsmath}
\usepackage{epsfig}
\usepackage{array}
\usepackage{multirow}
\usepackage{color}
\usepackage[english]{babel}
\usepackage{booktabs}
\usepackage{dcolumn}% Align table columns on decimal point
\usepackage{bm}% bold math
\usepackage{float}
\usepackage[
   bookmarks=true,
   colorlinks=true,
   hyperindex=true,
   citecolor=blue,
   linkcolor=blue,
]{hyperref}

\begin{document}

\title{Preparation of a Heteronuclear Two-atom System in the 3D Ground State in an Optical Tweezer}

\author{Kunpeng Wang}
\affiliation{State Key Laboratory of Magnetic Resonance and Atomic and Molecular Physics, Wuhan Institute of Physics and Mathematics, APM, Chinese Academy of Sciences, Wuhan 430071, China}
\affiliation{Center for Cold Atom Physics, Chinese Academy of Sciences, Wuhan 430071, China}
\affiliation{School of Physics, University of Chinese Academy of Sciences, Beijing 100049, China}

\author{Xiaodong He}
\email{hexd@wipm.ac.cn}
\affiliation{State Key Laboratory of Magnetic Resonance and Atomic and Molecular Physics, Wuhan Institute of Physics and Mathematics, APM, Chinese Academy of Sciences, Wuhan 430071, China}
\affiliation{Center for Cold Atom Physics, Chinese Academy of Sciences, Wuhan 430071, China}

\author{Ruijun Guo}
\affiliation{State Key Laboratory of Magnetic Resonance and Atomic and Molecular Physics, Wuhan Institute of Physics and Mathematics, APM, Chinese Academy of Sciences, Wuhan 430071, China}
\affiliation{Center for Cold Atom Physics, Chinese Academy of Sciences, Wuhan 430071, China}
\affiliation{School of Physics, University of Chinese Academy of Sciences, Beijing 100049, China}

\author{Peng Xu}
\affiliation{State Key Laboratory of Magnetic Resonance and Atomic and Molecular Physics, Wuhan Institute of Physics and Mathematics, APM, Chinese Academy of Sciences, Wuhan 430071, China}
\affiliation{Center for Cold Atom Physics, Chinese Academy of Sciences, Wuhan 430071, China}

\author{Cheng Sheng}
\affiliation{State Key Laboratory of Magnetic Resonance and Atomic and Molecular Physics, Wuhan Institute of Physics and Mathematics, APM, Chinese Academy of Sciences, Wuhan 430071, China}
\affiliation{Center for Cold Atom Physics, Chinese Academy of Sciences, Wuhan 430071, China}

\author{Jun Zhuang}
\affiliation{State Key Laboratory of Magnetic Resonance and Atomic and Molecular Physics, Wuhan Institute of Physics and Mathematics, APM, Chinese Academy of Sciences, Wuhan 430071, China}
\affiliation{Center for Cold Atom Physics, Chinese Academy of Sciences, Wuhan 430071, China}
\affiliation{School of Physics, University of Chinese Academy of Sciences, Beijing 100049, China}

\author{Zongyuan Xiong}
\affiliation{State Key Laboratory of Magnetic Resonance and Atomic and Molecular Physics, Wuhan Institute of Physics and Mathematics, APM, Chinese Academy of Sciences, Wuhan 430071, China}
\affiliation{Center for Cold Atom Physics, Chinese Academy of Sciences, Wuhan 430071, China}

\author{Min Liu}
\affiliation{State Key Laboratory of Magnetic Resonance and Atomic and Molecular Physics, Wuhan Institute of Physics and Mathematics, APM, Chinese Academy of Sciences, Wuhan 430071, China}
\affiliation{Center for Cold Atom Physics, Chinese Academy of Sciences, Wuhan 430071, China}

\author{Jin Wang}
\affiliation{State Key Laboratory of Magnetic Resonance and Atomic and Molecular Physics, Wuhan Institute of Physics and Mathematics, APM, Chinese Academy of Sciences, Wuhan 430071, China}
\affiliation{Center for Cold Atom Physics, Chinese Academy of Sciences, Wuhan 430071, China}

\author{Mingsheng Zhan}
\email{mszhan@wipm.ac.cn}
\affiliation{State Key Laboratory of Magnetic Resonance and Atomic and Molecular Physics, Wuhan Institute of Physics and Mathematics, APM, Chinese Academy of Sciences, Wuhan 430071, China}
\affiliation{Center for Cold Atom Physics, Chinese Academy of Sciences, Wuhan 430071, China}

\date{\today}

\begin{abstract}
We report the preparation of a heteronuclear two-atom system of $^{87}$Rb and $^{85}$Rb in the ground state in an optical tweezer. Dual-species Raman sideband cooling is applied to the two initially separated atoms to eliminate the crosstalk and a 3D ground-state probability of 0.91(5) for $^{87}$Rb and 0.91(10) for $^{85}$Rb are obtained. We then merge the  two atoms into one trap with a species-dependent transport which is achieved by utilizing vector light shifts depending on the magnetic moments of specific atomic states and the trap polarizations.
The measurable motional excitations due to merging are 0.013(1) and 0.006(3) axial vibrational quantums for the $^{87}$Rb and $^{85}$Rb atom respectively, while no obvious excitation is observed in the radial directions. This two-atom system offers a good starting point for building a single heteronuclear molecule and for investigating few-body physics. It can also be extended to other atomic species and molecules, and thus find application in ultracold chemistry.
\end{abstract}

%\pacs{03.67.-a, 42.50.Dv, 42.50.Ct}
%03.67.-a	Quantum information
%03.67.Lx	Quantum computation architectures and implementations
%42.50.Ct	Quantum description of interaction of light and matter; related experiments
%42.50.Dv	Quantum state engineering and measurements

\maketitle

\section{Introduction}
Isolated cold atoms or molecules have wide impacts in multiple disciplines ranging from collisional physics~\cite{Chin2010,Blume12RPP,Greene2017,PJulienneReview2019} and chemical reactions~\cite{KohlChemistry2012,Balakrishnan2016,JBohnReview2017} to quantum information and simulations~\cite{KaufmanEntanglement,twobodyJochim12,twobodyJochim15,YuanEntanglement2016,SCornishReview2018,BrowaeysTopological,LukinCat2019} and precision measurements \cite{ZelevinskyClock,KaufmanClock2019,EndresClock2019}. Compared with neutral atoms, cold molecules offer more rich internal states and  long-range interactions can be induced with electric fields.
Although two-atom pairs in optical lattices have been associated into single molecules \cite{BlochMolecule2004,DenschlagLattice2006,RempeLattice2006,BongsLattice2006}, bottom-up creation of single molecules in optical tweezers (OTs) offers new opportunities due to the flexible trap controllability and scalability, and moreover, the intrinsic features of single trap addressing and detection. Along this line, creation of an exited-state NaCs molecule from two atoms has been demonstrated recently by a single-photon association at finite atomic temperatures~\cite{Ni17Molecule}.
To produce single cold molecules with high efficiency, a crucial prerequisite is to prepare a two-atom system in the motional ground state of a single trap.
The technical breaking of inserting two atoms into a single trap has been demonstrated firstly with thermal Cs atoms in 2006 \cite{DieterInsert2006}, but the success rate of the inserting was only 16\% and neither ground state cooling nor preservation of motional states is implemented. After that, the combination of two heteronuclear atoms together into one tweezer was implemented with a success rate of about 95\% and  hyperfine-state dependent inelastic collisions between two atoms were observed \cite{NCSingleMolecule}, but again the atomic motional states were not controlled.

A reliable recipe for making an ultracold heteronuclear two-atom system is composed of two sequential steps: firstly, cool the two individual heteronuclear atoms to the 3D motional ground state, and then merge both the atoms into a single trap with negligible vibrational excitations. For the first step, the well developed Raman sideband cooling (RSC) technique~\cite{Monroe1995} can be deployed, as demonstrated with various homonuclear atoms in OTs from 1D to  3D~\cite{Kaufmancooling,Lukincooling,Dieter09,DWeiss12Cooling,Andersencooling,NiNacooling,Robens2018}. But a dual-species RSC of two heteronuclear atoms to the 3D ground state has not been reported yet. For the second step, to preserve the two-atom motional ground state,
it is essential to implement a species-dependent transport with an asymmetric potential configuration where each atom experiences a deeper local potential, far-off resonant with any vibrational levels of the other approaching one. With a species-dependent transport, the two atoms will be efficiently merged for interaction and molecular association, and split for atomic imaging when confirmation of molecules.
As demonstrated in optical lattices, by engineering vector light shifts (VLSs) single atoms can be transported over several lattice sites maintaining in the motional ground state \cite{Robens2018}, and two homonuclear atoms on different lattice sites can be brought into contact and then entangled~\cite{Mandel2003PRL,Mandel2003Nature,FollingSDL2018}. Up to date, the species-dependent transport for two heteronuclear single atoms has only been demonstrated with two thermal atoms \cite{Ni17Molecule}.

%%%%%%%%%%%%%%%%%%%%%%%%%%%%%%%%%%%%%%%%%%%%%%%%
\begin{figure*}[htb]
	\centering
	\includegraphics[width=14cm]{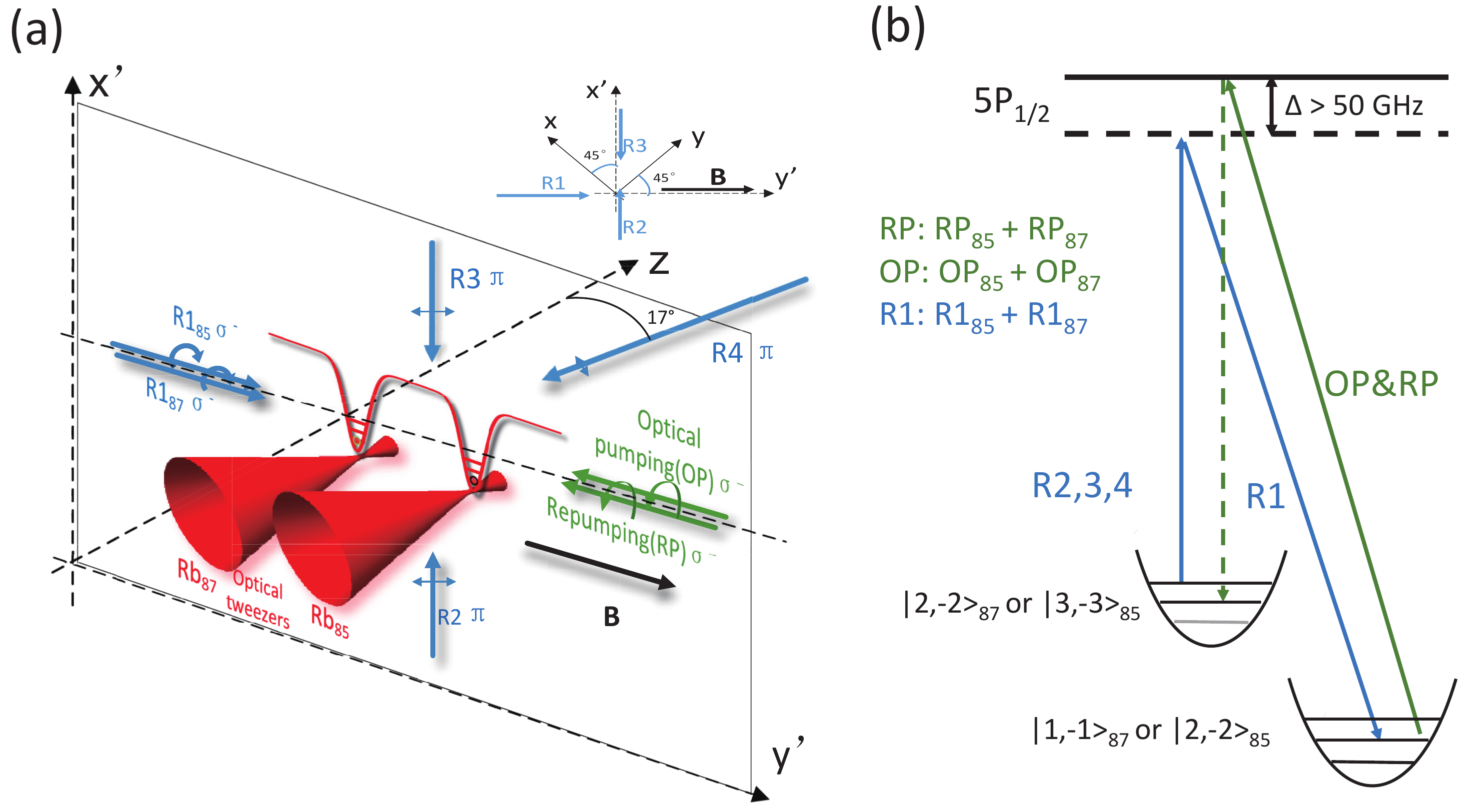}
	\caption{Experimental setup and Raman sideband cooling of heteronuclear $^{87}$Rb and $^{85}$Rb atoms. (a) Experimental arrangement. Two Raman R1 beams ($\sigma^{-}$ polarized), R1$_{87}$ for $^{87}$Rb and R1$_{85}$ for  $^{85}$Rb, are combined together with a beam splitter (BS) and propagate along the bias magnetic field direction y'. Optical pumping (OP,  $\sigma^{-}$) beams and repumping (RP, $\sigma^{-}$) beams for the two atoms are combined together with BSs before introduced into the glass cell.  R2 and R3 beams are approximately $\pi$ polarized to address the radial motions and propagate perpendicularly to the R1 beams. To address the axial motion of both atoms, R4 beam has a small angle with respect to the optical axis of trapping beam. (b) The simplified energy levels and transitions for the $^{87}$Rb and $^{85}$Rb atoms. The detail energy levels can be seen in Appendix \ref{APP-setup}. }
	\label{Experimentalsetup}
\end{figure*}
%%%%%%%%%%%%%%%%%%%%%%%%%%%%%%%%%%%%%%%%%%%%%%%%%%%%%%%%%%%%%%%%%%%%%%%%%%%

In this work, we demonstrate a general scheme to build a heteronuclear two-atom system in the ground state of a trap by utilizing a dual-species RSC and a species-dependent transport. We begin by simultaneous RSC of $^{87}$Rb and $^{85}$Rb in two traps and obtain a 3D ground-state with probability of 0.91(5) for $^{87}$Rb and 0.91(10) for $^{85}$Rb. And then merging the separated ground-state atoms into one trap is implemented with VLSs for specific spin states with opposite magnetic moments. The key point is to change the polarizations of the two OTs to the desired circular polarizations so that the state-dependent VLSs come into play. The measurable heating due to merging is below 0.02 vibrational quanta for both the two atoms in the axial dimension, while for the radial direction, no obvious heating is measured.

This paper is organized as follows. In Sec. \ref{sec_RSC}, we describe the experimental setup and the dual-species ground state cooling of $^{87}$Rb and $^{85}$Rb. In Sec. \ref{sec_mergingdynamics}, we analyze the single-atom merging dynamics in a symmetric double well.
In Sec. \ref{sec_SDtransport}, the scheme of species-dependent transport is proposed firstly, then the examination of this scheme by detecting qubit states and the two-atom merging experiments will be presented. In Sec. \ref{sec_conclusions}, we discuss the promising applications of the species-dependent transport and then conclude the paper.

\section{Dual-species ground state cooling}
\label{sec_RSC}

Figure \ref{Experimentalsetup} depicts the experimental setup. A static OT (S-trap) and a movable OT (M-trap) are used to trap one $^{85}$Rb atom and one $^{87}$Rb atom respectively
ensured by feedback controlled loading from a dual-species magneto-optical trap (MOT) \cite{NCSingleMolecule}. A high numerical aperture objective (NA = 0.6) is used to strongly focus two 852-nm trapping beams to waits of about 0.75 $\mu$m with a spacing of 4 $\mu$m. The spacing is controlled by a mirror actuated by piezo-electrical transducers
(PZTs) to insert the $^{87}$Rb atom from the M trap into the S trap.
For the typical trap depth of 1.6 mK, the radial and axial trap frequencies are $\omega_r = 2\pi \times $165 kHz and $\omega_z = 2\pi \times$ 27 kHz respectively. The two atoms are both further cooled to about 15 $\mu$K with standard optical molasses techniques. To avoid the polarization gradient effect, a static magnetic field of 6.7 G is set along y' direction ~\cite{Kaufmancooling,Lukincooling}.

The single-species RSC is typically implemented using four Raman beams to remove motional quanta and then optical pumps to initialize the atomic state and carries away entropy~\cite{Kaufmancooling,Lukincooling}. The total 3D RSC of an atom usually takes hundreds of milliseconds to achieve a 3D ground-state probability of above 90\%.
For the RSC of dual-species atoms, in our case $^{87}$Rb and $^{85}$Rb, the possible crosstalk between the cooling processes for the two atoms should be mitigated. By ``crosstalk'' we mean that RSC of the ``target'' atom would cause heating to the ``spectator'' atom by off-resonant photon scattering.
The near resonant pump (OP and RP) beams, large detuned Raman beams and the trap beam will cause heating of the spectator atom.
The pump beams will cause heating with a typical rate (in units of corresponding vibrational quanta per second, the same as below) of 0.18 (0.45)  for the radial (axial) dimension. The large detuned Raman beams, which are focused onto the atoms, cause an off-resonant heating rate of 0.8 (5) for each beam. And the trap beam will also heat the spectator atom with a typical rate of 0.6 (2.3)  in an 1.6-mK trap for the radial (axial) dimensions.
So the ground state probability of the spectator atom will be limited to about 30\% when the RSC are taken one-by-one for the two atoms. The crosstalk will also present in the Raman-sideband cooling of two atoms of different elements.

\begin{figure}[H]
	\centering
	\includegraphics[width=8.5cm]{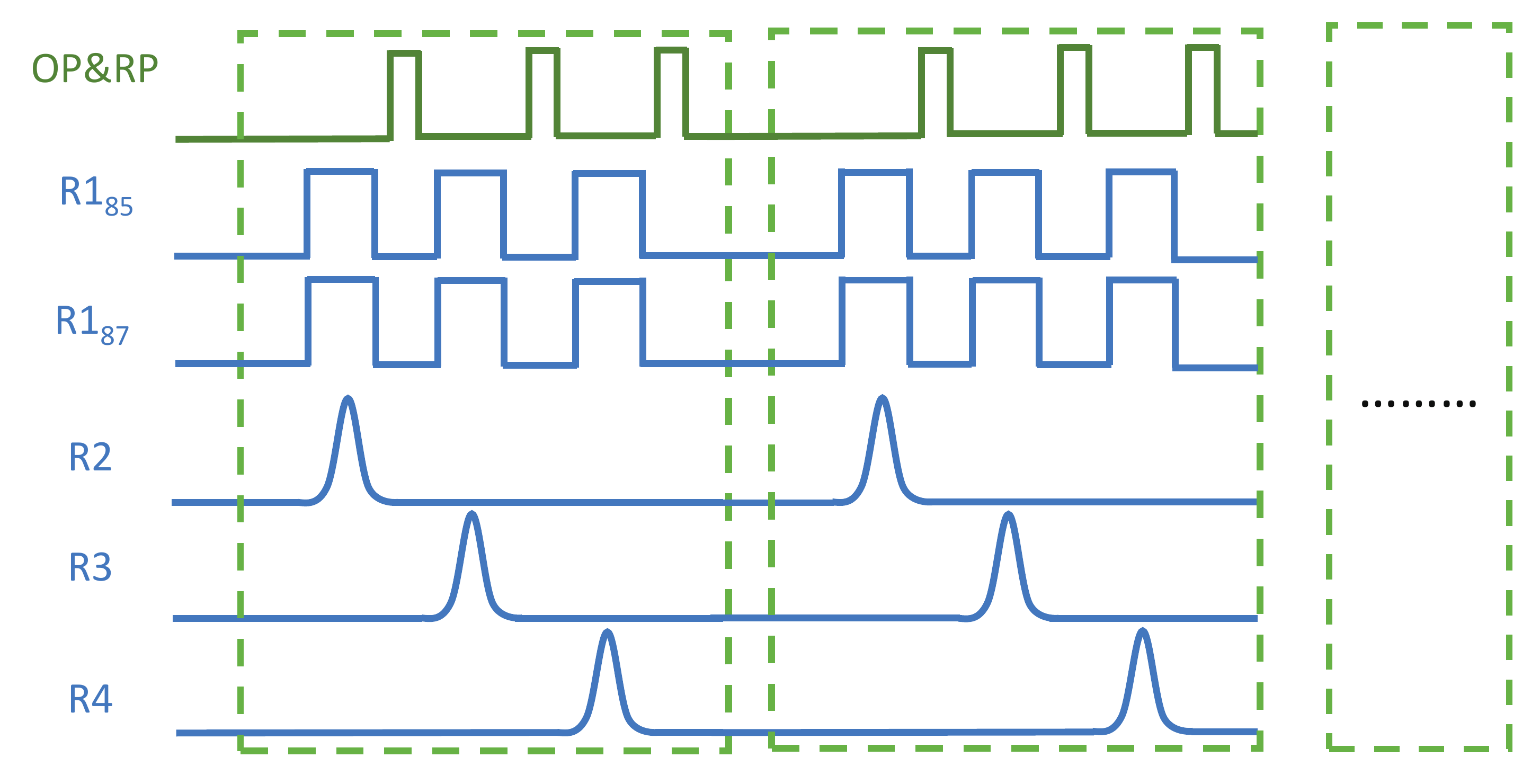}
	\caption{The dual-species RSC sequence. Each dashed square frame represents one RSC cycle.}
	\label{RSCsequence}
\end{figure}

To overcome the crosstalk, we design the dual-species RSC in an overlapped sequence with only five Raman beams as shown in Fig. \ref{RSCsequence}. The Raman laser setup are described in Appendix \ref{APP-setup}. While all the Raman and pump beams have large enough waists and cover both the two atoms, the  \{R2, R3, R4\} beams are shared for both the atoms. And the addressing of specific atoms is implemented with two spectral separated R1 beams, R1$_{85}$ and R1$_{87}$. During RSC the R1$_{85}$ and R1$_{87}$ beams are synchronously applied to both the two atoms to keep light shifts constant. And the resonant pump beams of $^{87}$Rb and $^{85}$Rb are also synchronously applied to the two atoms.
In this overlapped cooling sequence, the residual crosstalk is only induced by near resonant pump beams and the R1 beam of the target atom, which could heat the atoms by 0.001 (0.008) vibrational quanta for the radial (axial) direction in one typical cooling cycle.

\begin{figure}[H]
	\centering
	\includegraphics[width=8.6cm]{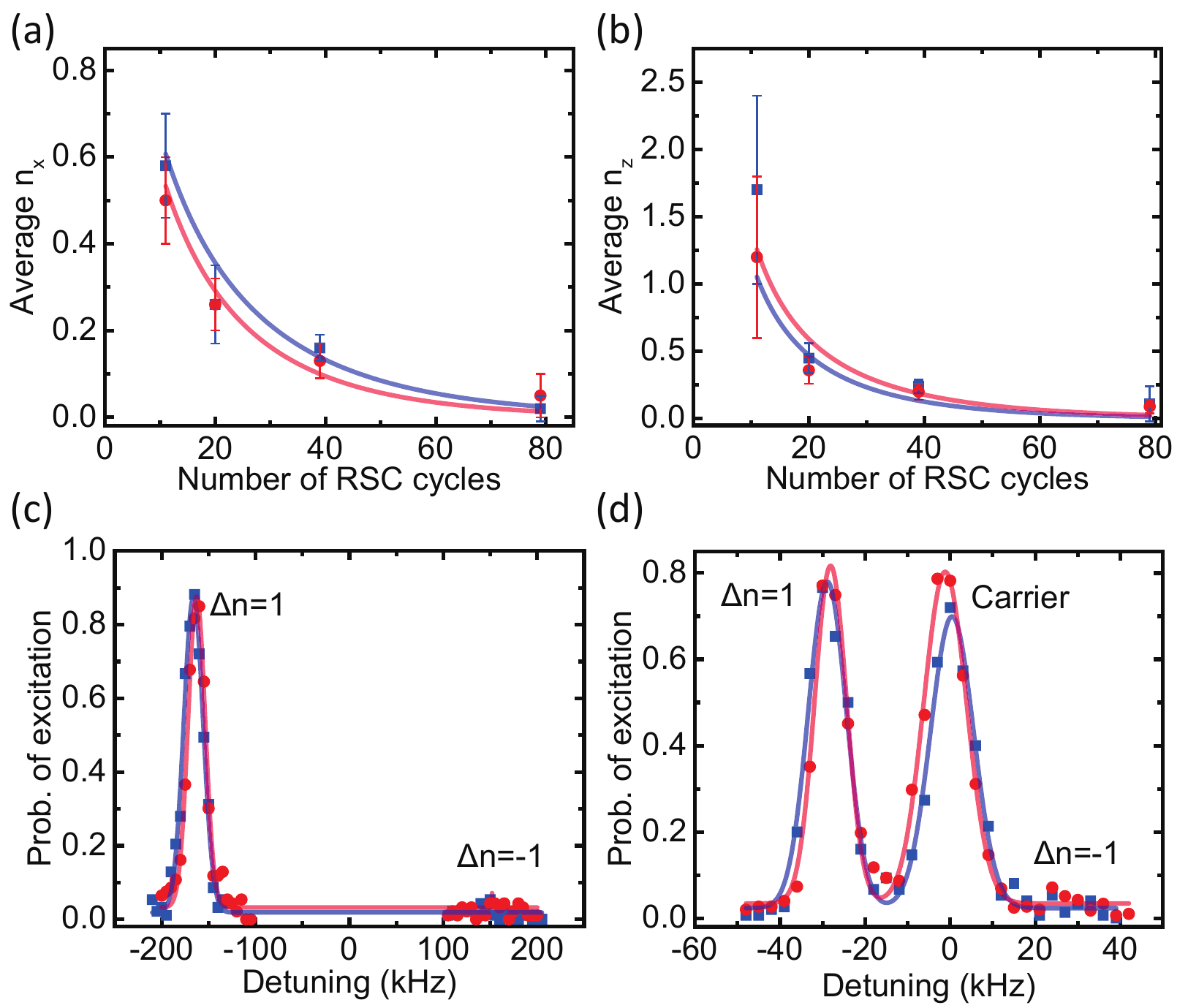}
	\caption{The measurement of the residual crosstalk and the Raman spectra after dual-species RSC. (a) and (b) show the measured $\overline{n}_{x,z}$ of $^{85}$Rb as a function of cooling cycles with (red circles) and without (blue squares) the participation of $^{87}$Rb cooling beams along x axis  and z axis respectively. The solid lines are fits to an exponential decay function.  (c) The radial RSC results of  $^{87}$Rb on the x (squares) and y (filled circles) dimension. (d) Axial dimensional cooling results of $^{87}$Rb (blue squares) and $^{85}$Rb (red circles).}
	\label{crosstalkandRSC}
\end{figure}

We examine the residual crosstalk by measuring the $\overline{n}_x$ and $\overline{n}_z$ of $^{85}$Rb after different numbers of cooling cycles with and without the participation of the cooling pulses of $^{87}$Rb, as shown in Fig. \ref{crosstalkandRSC}(a) and \ref{crosstalkandRSC}(b) respectively. The fluctuations in $\overline{n}$ is large, especially for a small number of RSC cycles. We fit the data points using a decay model of $\overline{n}=1/(exp(\hbar \omega(\alpha c+1)/k_B T)-1)$ \cite{Andersencooling}, where $c$ is the number of the cooling cycles and T is the initial atomic temperature before RSC. The fitted cooling rates of $^{85}$Rb for the radial dimension are $\alpha$ = 0.09(1) and 0.08(1) with and without the crosstalk from $^{87}$Rb respectively, resulting a difference of 0.01(1). For the axial dimension, the crosstalk lead to a difference of 0.08(8) in the cooling rates. Moreover, no obvious difference in $\overline{n}$ is observed after 80 RSC cycles for both the radial and axial dimensions. Thus the residual crosstalk can only have negligible influence on the final cooling fidelity.

The dual-species RSC contains 220 cooling cycles with a total duration of about 150 ms.  After cooling, the 3D motional quantum numbers $\{\overline{n}_{x},\overline{n}_{y},\overline{n}_{z}\}$ are  $\{0.04(3), 0.01(1), 0.04(4)\}$ for $^{87}$Rb and $\{0.01(4), 0.03(5), 0.05(9)\}$ for $^{85}$Rb. The final 3D ground-state probabilities are determined to be 0.91(5) for $^{87}$Rb and 0.91(10) for $^{85}$Rb respectively. The corresponding radial spectra  of $^{87}$Rb along x and y directions are shown in Fig. \ref{crosstalkandRSC}(c), and the axial spectra for both $^{87}$Rb and $^{85}$Rb are shown in Fig. \ref{crosstalkandRSC}(d).

%%%%%%%%%%%%%%%%%%%%%%%%%%%%%%%%%%%%%%%%%%%%%%%%
%%%%%%%%%%%%%%%%%%%%%%%%%%%%%%%%%%%%%%%%%%%%%%%%

\section{Merging dynamics of a symmetric double well}
\label{sec_mergingdynamics}

In this section, we study the merging dynamics of a single atom in a symmetric double-well potential and determine the heating mechanism during merging.

%%%%%%%%%%%%%%%%%%%%%%%%%%%%%%%%%%%%%%%%
\begin{figure}[htb]
	\centering
	\includegraphics[width=8.5cm]{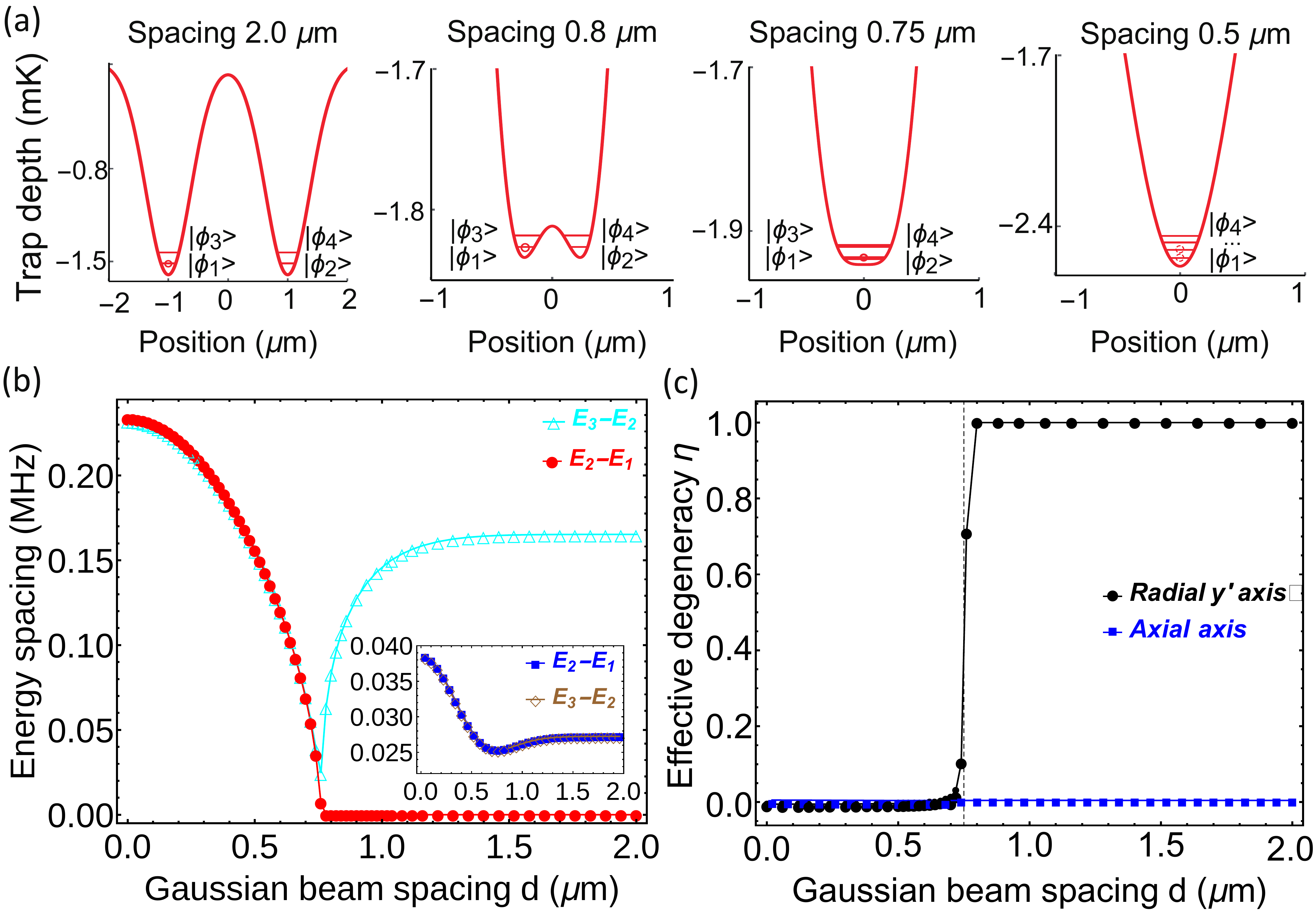}
	\caption{The merging dynamics of a symmetric double well with initial trap depths of 1.6 mK and trap-beam waists of 0.75 $\mu$m. (a) The snapshots for the merging process (energy levels not to scale). The spacing is the $d$ defined in Equ. \ref{equ_DW}. (b) The energy spacings for the radial x direction between the first ($|\phi_1\rangle$) and the second ($|\phi_2\rangle$) eigenstate are shown with filled circles, and the spacings between the second and the third ($|\phi_3\rangle$) eigenstate are shown with triangles. The inset shows the energy spacings of the axial direction. (c) The effective degree of degeneracy are shown as a function of trap beam spacings for the radial y' axis (filled circles) and axial axis (squares).}
	\label{SymmetricDWMerging}
\end{figure}
%%%%%%%%%%%%%%%%%%%%%%%%%%%%%%%%%%%%%%%%%%%%%%%%%%%%%%%%%%%%%%%%%%%%%

For two Gaussian traps, the potential can be described by
\begin{equation}
\label{equ_DW}
(V_0 + \Delta V) \exp \left(-\frac{2 \left(y' + d/2\right)^2}{\omega_0^2}\right)+V_0 \exp \left(-\frac{2 \left(y'-d/2\right)^2}{\omega_0^2}\right)
\end{equation}
, where $V_0$ and $V_0 + \Delta V$ are the trap depths, $d$ is the spacing of the beams center and $\omega_0$ is the beam waist.
A symmetrical double well, e.g. two identical Gaussian traps, corresponds to $\Delta V=0$. Typical merging snapshots at different trap spacings $d$ are illustrated in Fig. \ref{SymmetricDWMerging}(a). The merging axis is set along y' direction as defined in Fig. \ref{Experimentalsetup}(a). For the two traps that are totally separated before merging, two sets of harmonic oscillator eigenstates can be well defined as \{$|n=0\rangle_L$, $|n=1\rangle_L$, ...\} for the left well and \{$|n=0\rangle_R$, $|n=1\rangle_R$, ...\} for the right one. But when the two traps are fully overlapped, the system have only one set of harmonic oscillator eigenstates. To give a unified definition, the first four eigenstates of the system are redefined as $|\phi_1\rangle \equiv |n=0\rangle_L$, $|\phi_2\rangle \equiv |n=0\rangle_R$, $|\phi_3\rangle \equiv |n=1\rangle_L$ and $|\phi_4\rangle \equiv |n=1\rangle_R$. Before merging, the system have two-fold degenerate eigenstates i.e. the corresponding eigenenergy $E_{1}$ = $E_{2}$ and $E_{3}$ = $E_{4}$. To illustrate the character of degeneracy and its dynamical behavior during merging, we solve the eigenstates of the potential described in Eq. \ref{equ_DW} and
define an effective parameter $\eta$ for the ground state as
\begin{equation}
\label{Degeneracy}
\eta=1-\frac{E_{2}-E_{1}}{E_{3}-E_{2}}.
\end{equation}
The evolution of the energy spacings and the parameter $\eta$  are shown in Fig. \ref{SymmetricDWMerging}(b) and \ref{SymmetricDWMerging}(c) respectively. We note that the actual distance between the double-well potential minima is less than the spacing of the trap beams $d$, the same as below. During merging the system starts with a degenerate ground state which corresponds to a zero energy spacing between $|\phi_1\rangle$ and $|\phi_2\rangle$ and consequently $\eta=1$. As the trap spacing is decreased, the system crosses a critical point where the degeneracy is removed and then has a non-degenerate ground state, the corresponding spacings $E_{2}-E_{1}$ are equal to $E_{3}-E_{2}$ and thus $\eta=0$.

At the critical point, the atom exhibits dynamical instability and will undergo a ``quantum phase transition''. The general physics underlying this phenomenon is in analogy to the Jahn-Teller effect in polyatomic molecules, where the orbital instability of systems having degenerate states is pointed out \cite{Jahn1937}, and also in analogy to the fragmentation phenomenon in Bose-Eisenstein condensates \cite{CederbaumFragment2007,FischerFragment2009}.

Thus an atom initially in the ground state of one trap will dynamically evolve into the motional excited states with a finite probability of 50\% even for a slow enough merging process, which is confirmed with numerical integrations of the time-dependent Schr\"odinger equation, and the results will be presented in the next section.
We note that, in the region of $0<\eta<1$, the potential is anharmonic for the ground state.
As the traps are merged along the y' axis,  dynamics along the radial x' and axial axes are different, where the degeneracy parameters $\eta$ are always 0. The evolution of parameters for the axial dimension are shown in the inset of Fig. \ref{SymmetricDWMerging}(b) and squares in Fig. \ref{SymmetricDWMerging}(c).

%%%%%%%%%%%%%%%%%%%%%%%%%
\begin{figure}[H]
	\centering
	\includegraphics[width=8.5cm]{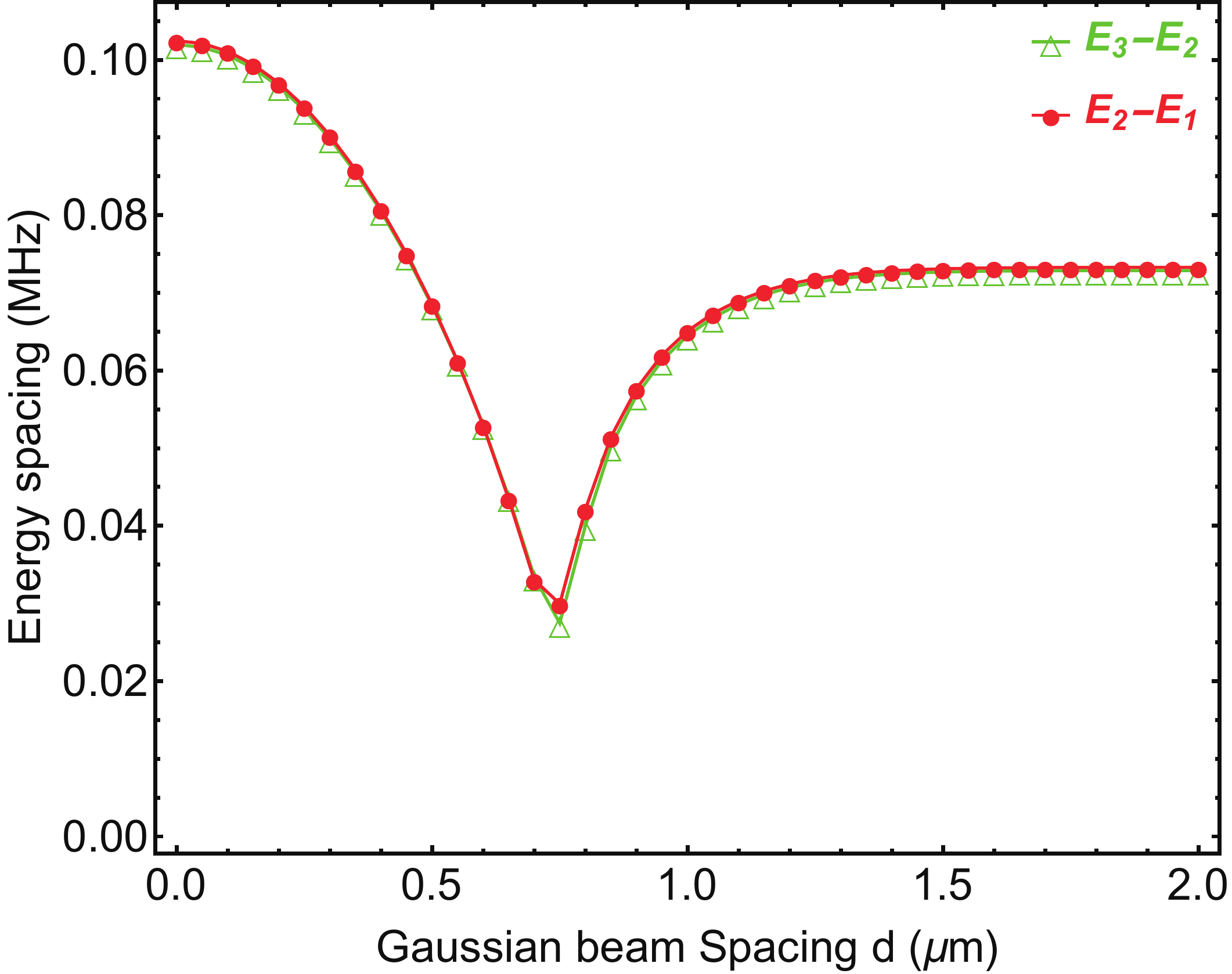}
	\caption{Merging dynamics of an asymmetric double well with depths of 0.3 mK and 0.315 mK respectively. The energy spacings between the first three eigenstates $E_{2}-E_{1}$ and $E_{3}-E_{2}$ are shown with filled circles and triangles respectively. }
	\label{ASmerging}
\end{figure}
%%%%%%%%%%%%%%%%%%%%%%%%%%%%%%%%%%%%%%%%%%%%%%

We give a further discussion on merging two atoms trapped in a symmetric double well here. As proposed in Ref. \cite{QNiu2003}, the two-atom motional ground state can be achieved by engineering the inter-atomic interaction using Feshbach resonances. In this scheme, starting with two atoms in a symmetric double well side-by-side, the atomic interaction should be tuned to be repulsive before merging and then tuned to be attractive  for splitting and return back to the original configuration. But this scheme is not suitable for searching for unknown Feshbach resonances and molecular association where two atoms need to be split into two traps for imaging. Moreover, in the setup with two separated traps, it is a challenge to construct an exactly symmetric double well due to relative intensity fluctuations. Thus we turn to an experimentally feasible and robust scheme for merging two atoms with an asymmetric double-well configuration, a species-dependent transport, to avoid ground-state degeneracy and mitigate trap fluctuations during merging and splitting.

\section{species-dependent transport}
\label{sec_SDtransport}

In this section, we analyze the dynamics of an asymmetric double well and the experimental conditions.
And we propose a robust scheme to construct a species-dependent transport using vector light shifts which can efficiently merge two atoms together for interaction and split them apart for imaging. We then study the merging process experimentally in the remaining part of this section.

\subsection{Proposal of species-dependent transport}
\label{subsec_SDT}
Typical merging dynamics of an asymmetric double-well configuration is studied as shown in Fig. \ref{ASmerging}, where one trap is set at $V_0 = $0.3 mK  and merged with a 5\% deeper trap ($\Delta V = 0.015$ mK). Although the trap frequency (energy spacing) at the critical spacing decreases, the degeneracy of the ground state is apparently removed. Thus in the asymmetric configuration, all the three dimensions have no degenerate ground states, and the 3D ground state can be preserved during an adiabatic merging process.

\begin{figure}[H]
	\centering
	\includegraphics[width=8.6cm]{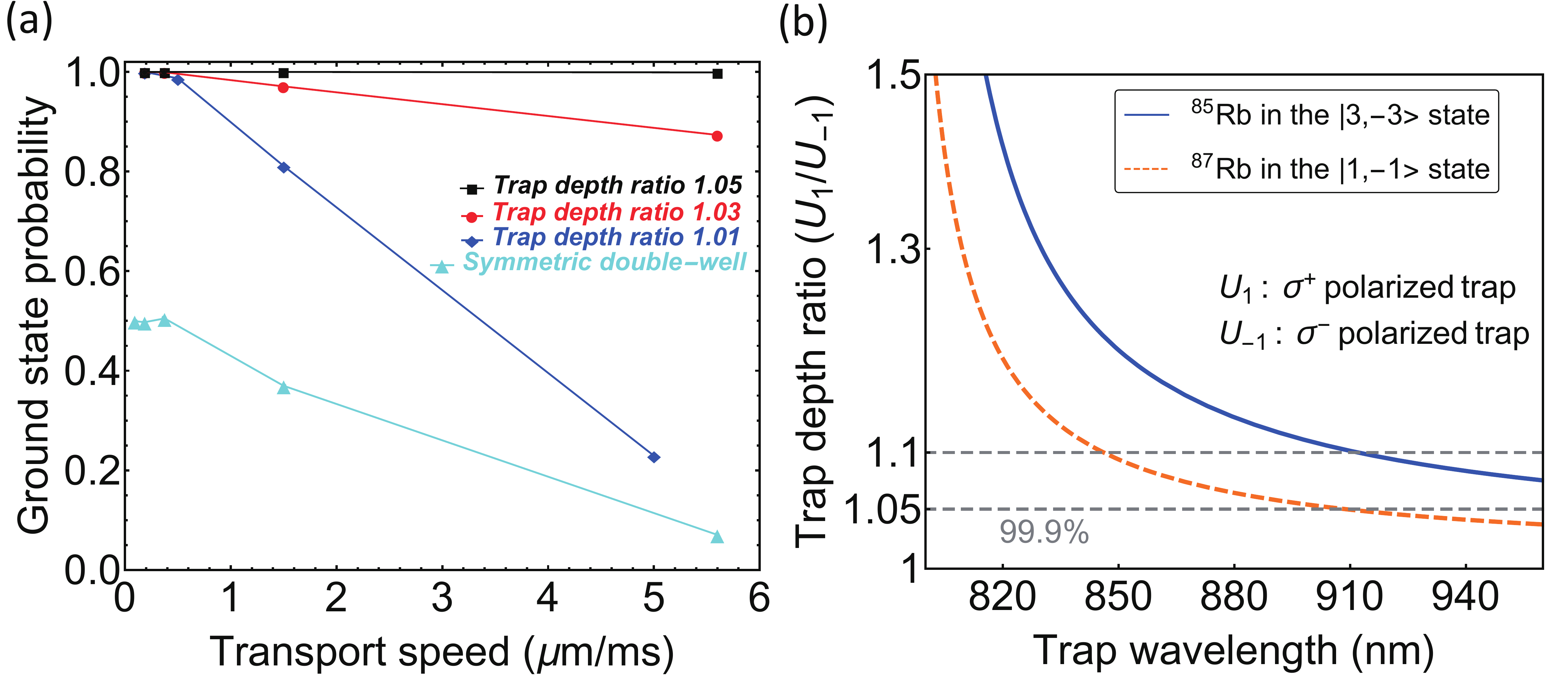}
	\caption{(a) The ground state probability along the merging axis after merging with different trap depth ratios and transport speeds. The triangles are the ground state probabilities for merging a symmetric double well. (b)  The trap-depth ratio of a $\sigma^+$ polarized trap (noted as U$_1$) and a  $\sigma^-$ polarized trap (U$_{-1}$) as a function of trap wavelength. The solid lines shows the ratio of the $^{85}$Rb atom in the $|3,-3\rangle$ state, and the dashed line shows the depth ratio of $^{87}$Rb atom in the $|1,-1\rangle$ state. }
	\label{mergingcondition}
\end{figure}

To guide the experiments, we numerically integrate the time-dependent Schr\"odinger equation for the transport process using the Crank-Nicolson method \cite{LatticeTransfer2008}. As shown in Fig. \ref{mergingcondition}(a), if one atom initially locates in a deeper trap and merged into a shallower trap with a depth ratio of 1.05, the atom will populate the motional ground state with a fidelity of 99.9\% for a wide range of transport speed of less than 5 $\mu$m/ms. For a lower trap-depth ratio, the transport speed should be lowered down to maintain a high ground state fidelity. However, for a symmetric double-well potential, the ground state fidelity will saturate at 50\% even with a sufficiently low transport speed.

If an asymmetric trap configuration can be constructed for both two atoms simultaneously, this special configuration can implement a ``species-dependent transport". For two atoms of different elements, this can be achieved by introducing two traps with different wavelengths utilizing the large difference in their scalar polarizabilities, as proposed and demonstrated in optical lattices or tweezers \cite{ThywissenSDLattice2007,GaaloulSDLattice2015,NagerlRbCs2017,Ni17Molecule}. For two isotopic atoms or two atoms in different spin states of a single species, they can be distinguished with VLSs at tune-out wavelengths \cite{Mandel2003PRL,Mandel2003Nature,FollingSDL2018,QCAEA2008}, however, tune-out lasers typically have small detunings from electronic resonant transitions for alkali atoms which could lead to large photon-scattering induced heating.

%%%%%%%%%%%%%%%%%%%%%%%%%%%%%%%%%%%%%%%%%%%%%%%%
%%%%%%%%%%%%%%%%%%%%%%%%%%%%%%%%%%%%%%%%%%%%%%%%

%%%%%%%%%%%%%%%%%%%%%%%%%%%%%%%%%%%%%%%%%%%%%%%%%%%%%%%%%%%%%%%%%%%%%%%
\begin{figure}[H]
	\centering
	\includegraphics[width=8cm]{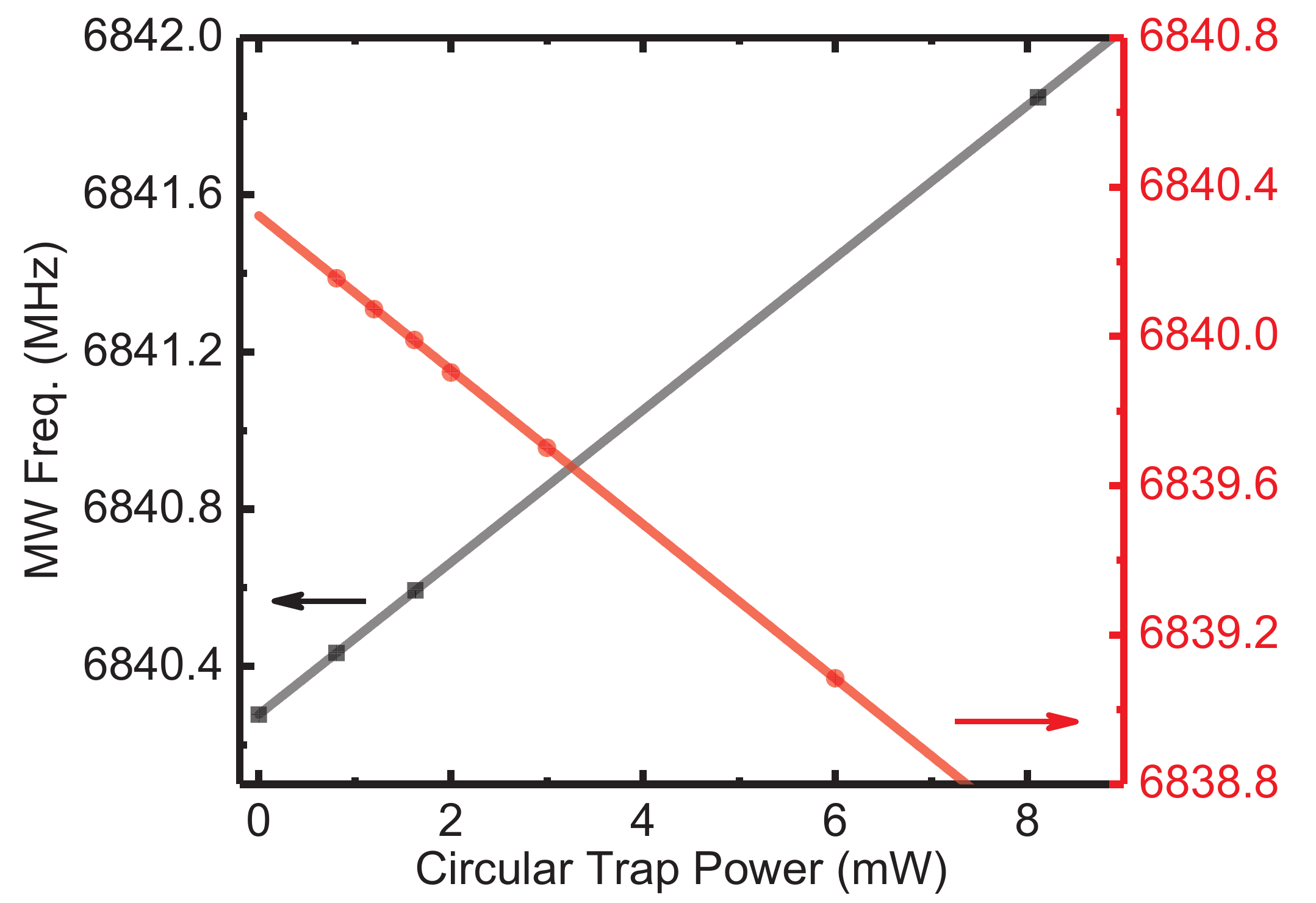}
	\caption{Measurements of the VLS with the MW transitions of $^{87}$Rb atoms from $|1,1\rangle$ to $|2,0\rangle$ state. The VLSs are only imparted for the $|1,1\rangle$ state while the $|2,0\rangle$ state only has tiny scalar shift. The MW resonant frequencies are shown as a function of optical power of circularly polarized S trap (black squares) and M trap (red dots). The fitted slope is 0.194 (-0.207) MHz/mW for S (M) trap.}
	\label{vectorlightshifts}
\end{figure}
%%%%%%%%%%%%%%%%%%%%%%%%%%%%%%%%%%%%%%%%%%%%%%%%%%%%%%%%%%%%%%%%%%%%%%%

Here we propose a robust scenario of species-dependent transport that utilizing VLSs with far-off resonant traps to merge and split two atoms in hyperfine states with opposite magnetic moments. In this scenario, two atoms with different internal states are respectively confined in $\sigma^+$ and  $\sigma^-$ polarized potentials in such a way that one atom dominantly experiences the $\sigma^+$ potential and the other mainly experiences the $\sigma^-$ potential. For two atomic states with opposite magnetic moments, such as $|3,-3\rangle_{85}$ (1.00 $\mu_B$ (Bohr magneton)) and $|1,-1\rangle_{87}$ ($-0.50 \, \mu_B$), their vector light shifts have opposite signs thus experience asymmetric trap configurations.

%%%%%%%%%%%%%%%%%%%%%%%%%%%%%%%%%%%%%%%%%%%%%%%%%%%%%%%%%%%%%%%%%%%%%
\begin{figure*}[htb]
	\centering
	\includegraphics[width=15cm]{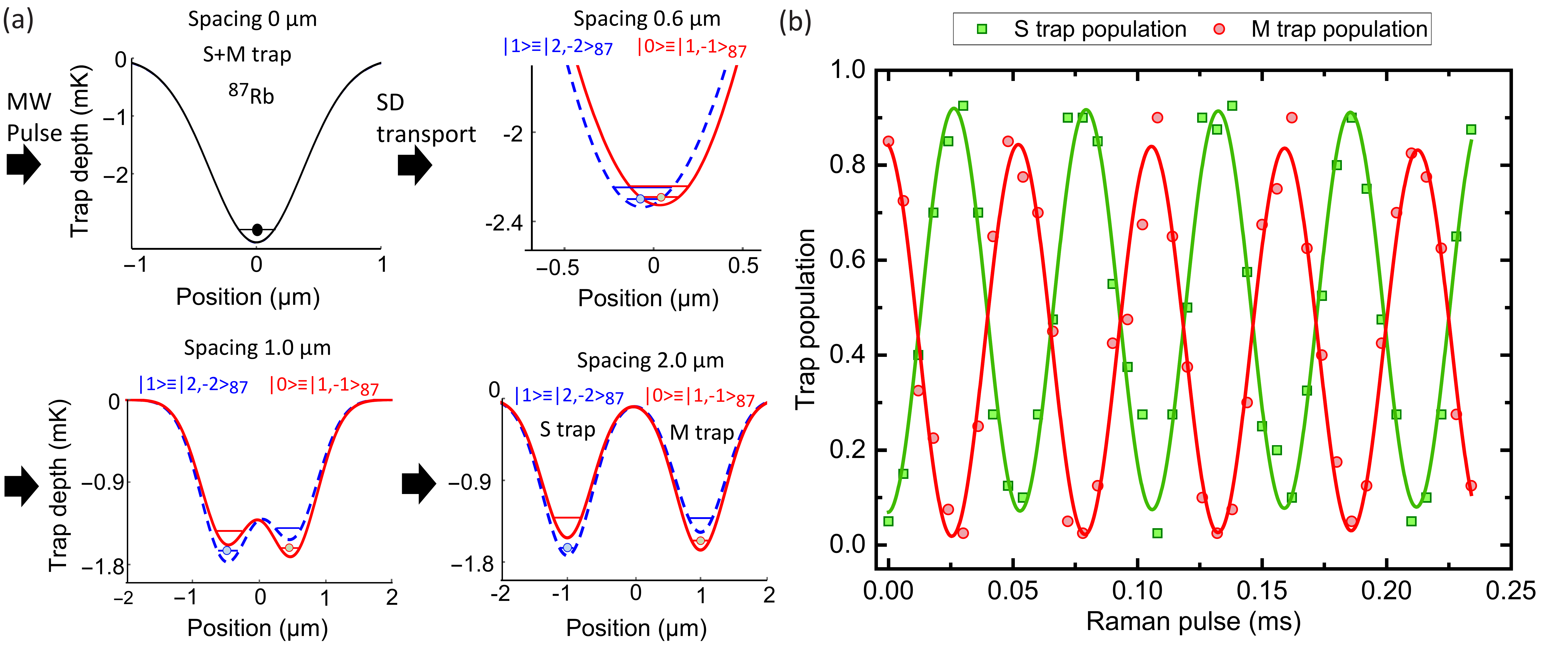}
	\caption{State detection of $^{87}$Rb between $|0\rangle\equiv|1,-1\rangle_{87}$ and $|1\rangle\equiv|2,-2\rangle_{87}$ using the VLS-dependent transport. (a) Snapshots of the detection process (energy levels not to scale) for spacings $d$ ranging from 0 to 4 $\mu$m. The dashed line and solid line are the potential for two atomic states that have opposite magnetic moments:  $|1\rangle$ and $|0\rangle$ respectively. (b) The oscillating behavior of probabilities in the S (M) trap are shown as squares (filled circles) corresponding to the probabilities of $^{87}$Rb in the $|1\rangle$ ($|0\rangle$) state. }
	\label{SDmerging}
\end{figure*}
%%%%%%%%%%%%%%%%%%%%%%%%%%%%%%%%%%%%%%%%%%%%%%%%%%%%%%%%%%%%%%%%%%%%%%
To confirm the experimental conditions for our system, we calculate the trap-depth ratios of $^{85}$Rb and $^{87}$Rb atom in a $\sigma^+$ polarized trap (denoted as $U_1$) and a  $\sigma^-$ polarized trap ($U_{-1}$) as shown in Fig. \ref{mergingcondition}(b). In our setup, we use an 852-nm laser for the two traps, the trap-depth ratios  $U_1/U_{-1}$ are larger than 1.09 for both the $|3,-3\rangle_{85}$ and $|1,-1\rangle_{87}$ state, which are well above the required value of 1.05 even when including the relative fluctuations in trap intensities of about 0.5\%.

The actual VLSs in circularly polarized traps are measured via microwave spectra of $^{87}$Rb.
To induce the VLSs, the trap polarizations are switched to right circularly polarized ($\sigma^{+}$) for S trap and left circularly polarized ($\sigma^{-}$) for M trap, and the magnetic field is switched from the y' to the z direction parallel to the trap beams.
The corresponding ideal normalized Stokes parameters are ($S_1, S_2, S_3$)=(0, 0, 1) for S trap and (0, 0, -1) for M trap respectively.
As shown in Fig. \ref{vectorlightshifts}, the MW resonant frequencies of $^{87}$Rb atoms decrease with the increasing optical power due to the VLS. The ratio of the fitted slope of the VLSs between S and M trap is 0.94, which is lower than 1 because of imperfect polarization settings.
Thus when initial depths of M and S traps are set at 0.3 mK, the actual trap depths differs by $2\pi \times (-0.56)$MHz ($2\pi \times 1.12 $MHz) for $^{87}$Rb ($^{85}$Rb) in the $|1,-1\rangle_{87}$ ($|3,-3\rangle_{85}$) state, which is much larger than the corresponding radial trap frequency of about $2\pi \times 72 $ kHz.

\subsection{Qubit state detection with a species-dependent transport}
\label{sec_SDdetection}

Before the two-atom merging experiment, we first examine the species-selectivity with the $^{87}$Rb atom in two spin states of $|0\rangle\equiv|1,-1\rangle_{87}$ and $|1\rangle\equiv|2,-2\rangle_{87}$. For the $|1\rangle$ state, it has a same magnetic moment of 1.00 $\mu_B$ as $|3,-3\rangle_{85}$, thus they will experience similar transport dynamics.
As shown in Fig. \ref{SDmerging}(a), the $^{87}$Rb atom is initially prepared in the $|0\rangle$ state and trapped in the 1.6-mK M trap, then the Raman R2 carrier transition are driven with various pulse durations to prepare a spin superposition states. Then the trap beams are set to the desired polarizations and the magnetic field are set to the z direction as described in Sec. \ref{subsec_SDT}.
During the polarization change process, the S trap ($\sigma^{+}$), initially overlapped with M trap ($d=0$ $\mu$m), is adiabatically turned on. Under this condition, the correlations between the two spin states and the two traps are established in the form of $|0,M\rangle$ and $|1,S\rangle$. By transporting the M trap apart, the $^{87}$Rb atom will locate in the M trap when in the $|0\rangle$ state and locate in the S trap when in the $|1\rangle$ state. The final probabilities of the atom in the two traps relate to the probabilities of the two spin states. The coherent Rabi oscillations between the two spin states $|0\rangle$ and $|1\rangle$ are thus detected as an oscillating behavior of survival probabilities in the two traps, as shown in Fig. \ref{SDmerging}(b). The finite amplitudes are limited by single-atom loss and initial state preparation fidelity and Raman transfer fidelity. The single-atom loss is about 4\% which is limited by the finite atomic lifetime of about 7 seconds. A high detection fidelity of about 95\% is achieved by benchmarking with the push-out detection technique. This powerful state detection technique that utilizing the VLSs has the potential to detect a single qubit among a large atoms array \cite{PortoLattice2007,Meschede2017,DWeissSD2018}.

\subsection{Two-atom merging experiments}
\label{sec_merging}

%%%%%%%%%%%%%%%%%%%%%%%%%%%%%%%%%%%%%%%%%%%%%%%%%%%%%%%%%%%%%%%%%%%%%%%%%%%%%%
\begin{figure*}[htb]
	\centering
	\includegraphics[width=14cm]{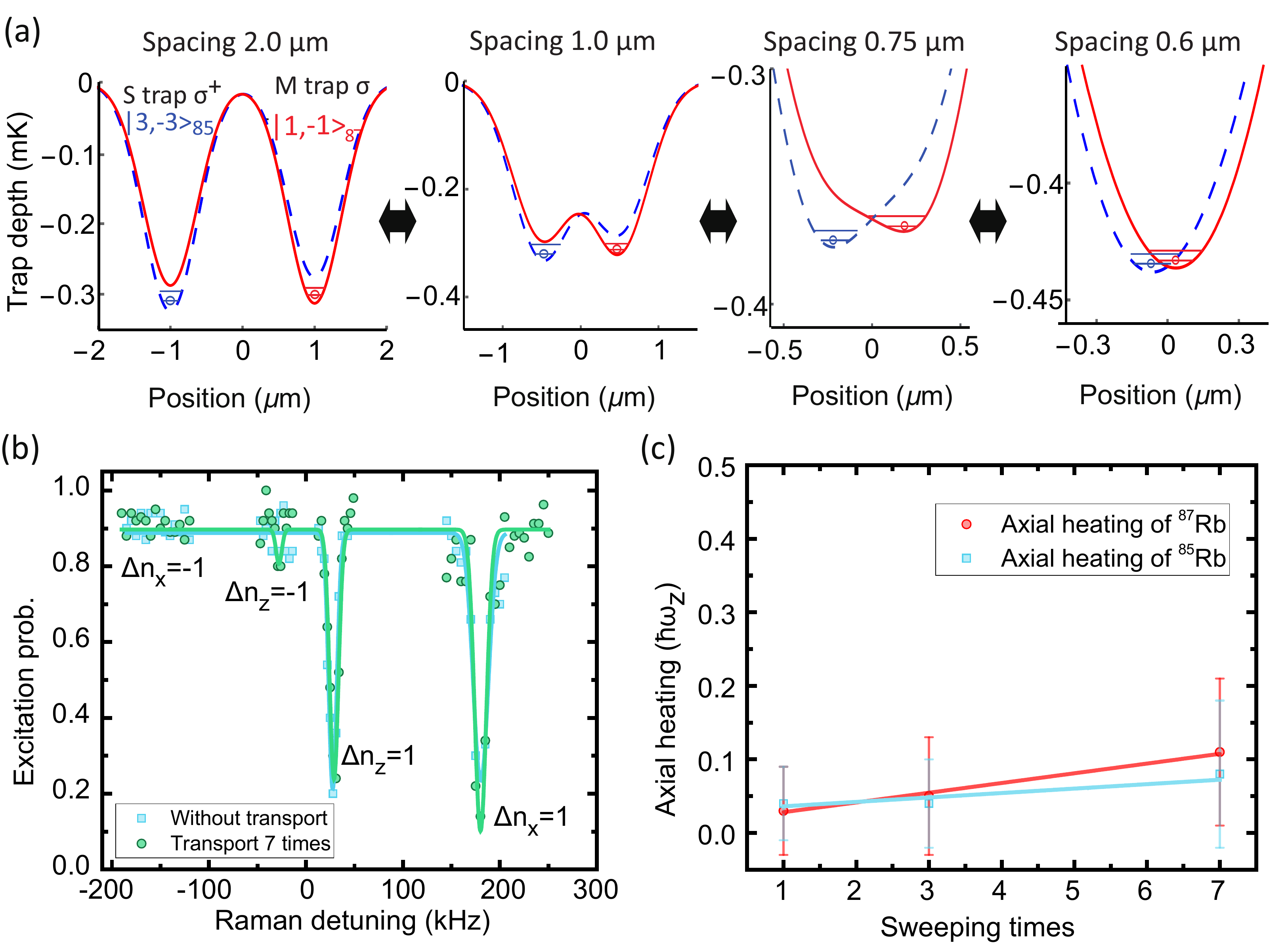}
	\caption{The illustration of the two-atom merging process and the measurement of heating due to merging. (a) Snapshots of the species-dependent merging process for spacings $d$ ranging from 4 to 0 $\mu$m (energy levels not to scale). The potential for two atomic states of $|3,-3\rangle_{85}$ and $|1,-1\rangle_{87}$ are shown with the dashed and solid lines respectively. (b) The filled circles show the sideband spectrum after repeating the merging and splitting process for 7 times, while the squares are of holding the $^{87}$Rb atoms for the same duration but without moving the M-trap. The heating for sweeping the PZT 7 times are extracted as the increment of $\overline{n}$. Note that the initial internal state is $|1,-1\rangle$, so the $\Delta$n=1 peaks have positive detunings compared with the spectra in Fig. \ref{Experimentalsetup}(c) and (d). (c) The data points show the extracted axial heating in $\overline{n}$ of $^{85}$Rb (squares) and $^{87}$Rb (filled circles) for sweeping the PZT 1, 3 and 7 times, where the accompanying error bars are obtained from the fit amplitude of the spectra. The average heating of the two atoms for one sweeping process are extracted as the slope of linear fit.}
	\label{mergingresults}
\end{figure*}
%%%%%%%%%%%%%%%%%%%%%%%%%%%%%%%%%%%%%%%%%%%%%%%%%%%%%%%%%%%%%%%%%%%%%%%%%%%%

For the two-atom merging of $^{87}$Rb and $^{85}$Rb atoms, we choose two spin-stretched states of $|3,-3\rangle_{85}$ and $|1,-1\rangle_{87}$, for which the inelastic collisions are energetically forbidden. The transport processes are shown in Fig. \ref{mergingresults}(a) and the detail experimental sequences are described in the Appendix \ref{APP-transport}. To lower down the heating from photon scattering of traps, the  intensity of S trap is lowered down (corresponding to 0.3 mK for a linearly polarized trap) during transport and the M trap is set according to the measured VLS ratio of the two traps.
Such a potential configuration makes each atom experience a deeper local potential and far-off resonant with any vibrational levels of the other approaching one. The trap polarizations and the constant magnetic field are the same as the case of state-detection described in last section.

The merging is implemented within 4 ms by transporting the $^{87}$Rb atoms in the M trap adiabatically to the position of the S trap. To lower down the transport speed at the critical region, the transport duration for the last 1-$\mu$m is set to 2 ms. Thereafter, the M-trap is adiabatically ramped off and the trap polarization and the magnetic field are switched back to the original values used during RSC. To measure the heating during merging, we repeat the merging and reversely splitting process by sweeping the PZT back and forth several times.
The sideband spectra with (filled circles) and without (squares) sweeping the PZT for 7 times are shown in Fig. \ref{mergingresults}(b). The heating due to sweeping is defined by the increment in $\overline{n}$. The dependencies of axial heating on the sweeping times are shown in Fig. \ref{mergingresults}(c). The heating per sweep for the axial motion is estimated to be 0.013(1) for the $^{87}$Rb atom (filled circles) and 0.006(3) for $^{85}$Rb (squares). For the radial motions, no obvious heating is measured after sweeping 7 times.

While the heating during merging is very small, the final ground-state probability for two atoms in one trap is limited by the initial RSC fidelity and additional heating caused by the photon scattering and intensity noise of the trap beams. The 3D ground-state probability after merging is estimated to be about 0.75 for one atom, leading to a probability of 0.56 for two atoms. The RSC can be further optimized by using high order Raman transitions \cite{NiNacooling} or by using a newly proposed post-selection scheme
\cite{CWZhang2019} and the background heating can be reduced by using a larger detuned trap and a fast transport \cite{MugaTransport2011}. For a larger detuned trap beam with a wavelength of 900 nm the species-dependent transport  can still be performed efficiently.

\section{conclusion}
\label{sec_conclusions}

In conclusion, we have demonstrated a general and robust scheme to prepare a low entropy atom pair with RSC and a species-dependent transport using an ultracold  pair of $^{87}$Rb and $^{85}$Rb atoms in the motional ground state of an OT. 

The species-dependent transport utilizing the VLSs demonstrated here can be extended to other systems with large magnetic moments such as rare-earth atoms (7 $\mu_B$ for Cr and Er, 10 $\mu_B$ for Dy)
\cite{PfauCrBEC2005,LevDyBEC2011,FerlainoEr2012} and even molecules. Based on the recent breakthrough of isolating single CaF molecules in OTs \cite{JDoyleArray19}, we predict that our method can also be applied to merge two CaF molecules in the ground rotational  state $|N=0, F=1\rangle$. This state has a large magnetic moment of about 1.0 $\mu_B$, which can induce a large trapping potential difference for the $|N=0, F=1, m_F=\pm1\rangle$ states using vector light shifts.

We also expect that such a low-entropy heteronuclear two-atom system is a promising starting-point for associating a single molecule with Feshbach resonances ~\cite{YouLiFR}. Furthermore, the two individual atoms in the 3D ground state have the promising potential to improve the Rydberg state mediated gate fidelity and entanglement fidelity~\cite{yz,BrowaeysEntanglement2010,SaffmanEntanglement2019,LukinEntanglement2019}. Our methods demonstrated here can be scaled up to few-atom regimes and find applications in studying three-body and few-body physics with various interesting systems~\cite{Blume2018,Blume2019,Kolck2019}, or specifically the equal-mass systems \cite{ParishImpurity2018}.

\textit{Note added} - We became aware of the demonstration of ground state cooling and merging of single Na and Cs atoms \cite{NiMolecule2019} after submission of this manuscript.
%%%%%%%%%%%%%%%%%%%%%%%%%%%

\begin{acknowledgments}
The authors would like to thank Yiheng Lin  for fruitful discussions and Yuan Gao for technical inputs. K.-P. Wang acknowledges special thanks to K.-K. Ni for discussions during ICAP2018.
This work was supported by the National Key Research and Development Program of China under Grant No. 2017YFA0304501, No. 2016YFA0302800, and No. 2016YFA0302002, the National Natural Science Foundation of China under Grant No. 11774389, the Strategic Priority Research Program of the Chinese Academy of Sciences under Grant No. XDB21010100, and the Youth Innovation Promotion Association CAS No. 2019325.
\end{acknowledgments}

\appendix
\setcounter{figure}{0}
\renewcommand\thefigure{A\arabic{figure}}

\section{Experimental details of the dual-species RSC}
\label{APP-setup}

\begin{figure}[H]
	\centering
	\includegraphics[width=8.6cm]{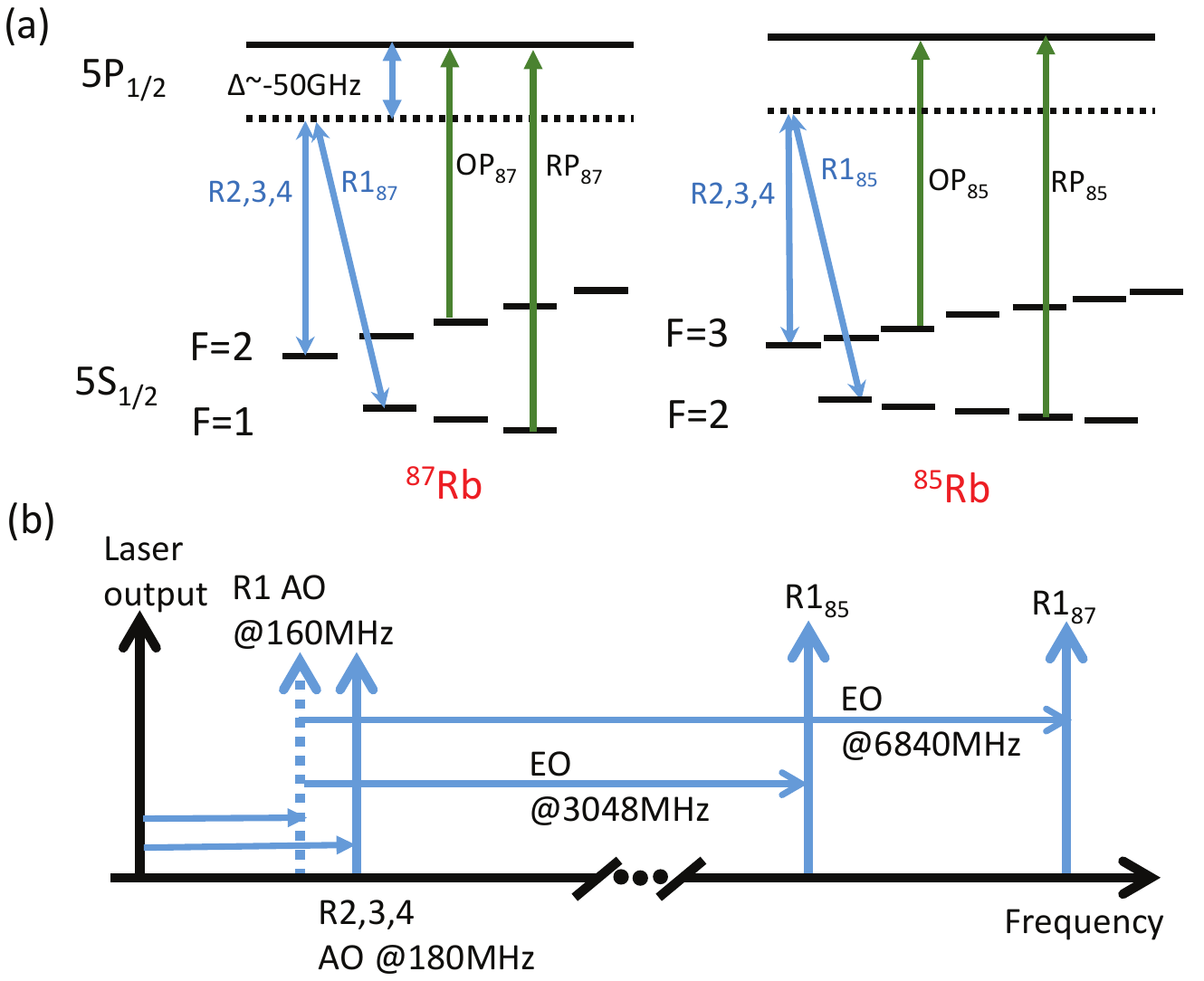}
	\caption{(a) The related energy levels and transitions for the $^{87}$Rb and $^{85}$Rb. (b) Generation of Raman beams.}
	\label{Ramanbeams}
\end{figure}

The related energy levels of RSC are shown in Fig. \ref{Ramanbeams}(a). The OP$_{87}$ beam is resonant with $5S_{1/2} F=2\rightarrow 5P_{1/2} F'=2$  and RP$_{87}$ is resonant with $5S_{1/2} F=1\rightarrow 5P_{1/2} F'=2$  for $^{87}$Rb atoms in the presence of bias magnetic field. The OP$_{85}$ beam is resonant with  $5S_{1/2} F=3\rightarrow 5P_{1/2} F'=3$ and RP$_{85}$ is resonant with $5S_{1/2} F=2\rightarrow 5P_{1/2} F'=3$ for $^{85}$Rb. The \{R2, R3, R4\} beams couples $|2,-2\rangle_{87}$ ($|3,-3\rangle_{85}$) and $5P_{1/2}$ states for $^{87}$Rb ($^{85}$Rb) and R1$_{87}$ (R1$_{85}$) couples the $|1,-1\rangle_{87}$ ($|2,-2\rangle_{85}$) and $5P_{1/2}$ state.

%%%%%%%%%%%%%%%%%%%%%%%%%%%%%%%%%%%%%%

%%%%%%%%%%%%%%%%%%%%%%%%%%%%%%%%%%%%%%%%%%%%%%%%%%%%%%%%%%%%%%%%%%%%%%%

%%%%%%%%%%%%%%%%%%%%%%%%%%%%%%%%%%%%%%%%%%%%%%%%%%%%%%%%%%%%%%%%%%%%%%%
\begin{figure}[H]
	\centering
	\includegraphics[width=8.6cm]{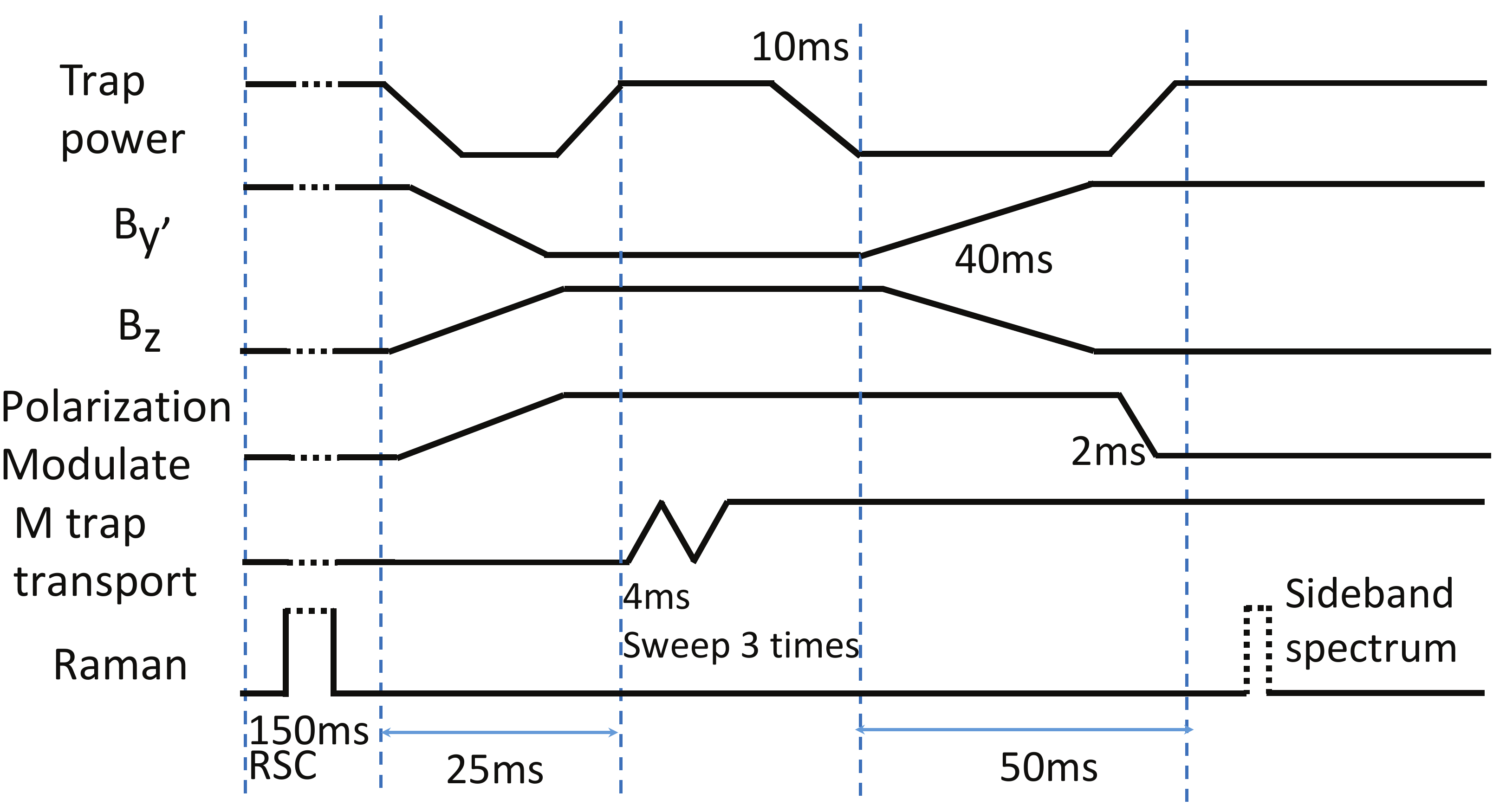}
	\caption{The experimental sequence of sweeping trap for 3 times.}
	\label{mergingsequence}
\end{figure}
%%%%%%%%%%%%%%%%%%%%%%%%%%%%%%%%%%%%%%%%%%%%%%%%%%%%%%%%%%%%%%%%%%%%%%%

To maintain the laser coherence and phases, all the Raman beams are derived from a single diode laser which is -50 GHz detuned of  $^{87}$Rb D1 line, and then frequency shifted with acousto-optic (AO) modulators as shown in Fig. \ref{Ramanbeams}(b). The R1$_{87}$ (R1$_{85}$) beam is then passes through an electro-optic (EO) phase modulator, operating at 6.832 GHz (3.035 GHz) to couple $|1,-1\rangle$ ($|2,-2\rangle$) and $5P_{1/2}$ for $^{87}$Rb ($^{85}$Rb) with the +1st order EO sideband. The optical powers of the each beams are actively locked after fiber couplers with drifts of less than 0.005 within 1 hour using analog Proportional Integral (PI) regulators. The R1 beam for $^{87}$Rb is then focused to a waist of 80 $\mu$m. And the other four beams are focused to waists of 110 $\mu$m. To avoid carrier heating during sideband cooling, the intensity profiles of \{R2, R3, R4\} beams are shaped to Gaussian pulses with calibrated waveforms stored in arbitrary waveform generators. The R1 beams typically induce a differential light shifts of about 40 kHz for the Raman transitions, so they are shaped as rectangular pulses to keep the light shifts constant.  Additionally, we note that the \{R2, R3, R4\} beams also cause light shifts varying with the Gaussian profiles, which are not compensated during RSC in this experiment. But the compensation is needed to drive high fidelity Raman sideband transitions.

\section{Experimental sequence of the transport process}
\label{APP-transport}

Fig. \ref{mergingsequence} shows the experimental sequence for transport 3 times. After merging, the trap polarization and the magnetic field are switched back to the original values used during RSC. This process costs 50 ms to stabilize the magnetic field. And then the Raman pulses are introduced. To measure the averaging heating during merging, we repeat the transport process for 1, 3, and 7 times and extract the slope of heating. The obtained sideband spectrum and heating are shown in Fig. \ref{mergingresults}(b) and \ref{mergingresults}(c) in the main text.

\bibliographystyle{apsrev4-1}

\end{document}